\documentclass[prl,showpacs,twocolumn,preprintnumbers]{revtex4}
\usepackage{graphicx}

\newcommand{\fref}[1]{Fig.~\ref{#1}}
\newcommand{\deff}{d_{\rm eff}}

\newcommand{\PZT}{PbZr$_{0.2}$Ti$_{0.8}$O$_3$~}
\newcommand{\PTO}{PbTiO$_3$~}

\begin{document}
\title{Domain wall roughness in epitaxial ferroelectric PbZr$_{0.2}$Ti$_{0.8}$O$_3$ thin films}
\author{P. Paruch}
%\email{}
\author{T. Giamarchi}
%\email{}
\author{J.-M. Triscone}
%\email{}
\affiliation{DPMC, University of Geneva, 24 Quai E. Ansermet, 1211
Geneva 4, Switzerland}
\date{\today}
\begin{abstract}
The static configuration of ferroelectric domain walls was
investigated  using atomic force microscopy on epitaxial
PbZr$_{0.2}$Ti$_{0.8}$O$_3$ thin films. Measurements of domain
wall roughness reveal a power law growth of the correlation
function of relative displacements $B(L) \propto L^{2\zeta}$ with
$\zeta \sim 0.26$ at short length scales $L$, followed by an
apparent saturation at large $L$. In the same films, the dynamic
exponent $\mu$ was found to be $\sim 0.6$ from independent
measurements of domain wall creep.  These results give an
effective domain wall dimensionality of $d=2.5$, in good agreement
with theoretical calculations for a two-dimensional elastic
interface in the presence of random-bond disorder and long range
dipolar interactions.
\end{abstract}
\pacs{77.80.-e, 77.80.Dj, 77.80.Fm}
\maketitle

Understanding the behavior of elastic objects pinned by periodic
or disorder potentials is of crucial importance for a large number
of physical systems ranging from vortex lattices in type II
superconductors \cite{blatter_vortex_review}, charge density waves
\cite{gruner_revue_cdw} and Wigner crystals
\cite{andrei_wigner_2d} to interfaces during growth
\cite{kardar_review_lines} and fluid invasion
\cite{wilkinson_invasion} processes, and magnetic domain walls
\cite{lemerle_FMDW}.  Ferroelectric materials, whose switchable
polarization and piezoelectric and pyroelectric properties make
them particularly promising for applications such as non-volatile
memories \cite{scott_memories,waser_memories}, actuators, and
sensors \cite{caliendo_saw_sensor}, are another such system. In
these materials, regions with different symmetry-equivalent ground
states characterized by a stable remanent polarization are
separated by elastic domain walls. The application of an electric
field favors one polarization state by reducing the energy
necessary to create a nucleus with polarization parallel to the
field, and thereby promotes domain wall motion. Since most of the
proposed applications use multi-domain configurations,
understanding the mechanisms that control domain wall propagation
and pinning in ferroelectrics is an important issue.

A phenomenological model derived from measurements of domain
growth in bulk ferroelectrics
\cite{merz_nonlinear_FE,fatuzzo_nonlinear_FE,miller_nucleation_FE}
initially suggested that the domain walls were pinned by the
periodic potential of the crystal lattice itself. Such pinning was
deemed possible because of the extreme thinness of ferroelectric
domain walls (different from the case of magnetic systems).
However, measurements of the piezoelectric effect
\cite{damjanovic_piezo}, dielectric permittivity
\cite{taylor_dielectric}, and dielectric dispersion
\cite{mueller_DW_pinning} in ferroelectric ceramics and sol-gel
films have shown some features which cannot be described by the
existing phenomenological theories. A microscopic study of
ferroelectric domain walls could resolve these issues. Recently,
we have measured domain wall velocity in epitaxial \PZT thin
films, showing that in this case commensurate lattice pining is in
fact not the dominant mechanism
\cite{tybell_creep,PP_dw_dynamics_FE}. Rather, a creep-like
velocity ($v$) response to an externally applied electric field
$E$ was observed with $v \sim \exp[-C/E^\mu]$, where $C$ is a
constant. The exponent $\mu$ characterizing the dynamic behavior
of the system is a function of the domain wall dimensionality and
the nature of the pinning potential. These results suggested that
domain wall creep in ferroelectric films is a disorder-controlled
process. However, questions about the microscopic nature of the
disorder were left open by the dynamical measurements alone. In
order to ascertain the precise physics of the pinned domain walls
and also the possible role of the long-range dipolar interactions
which exist in ferroelectric materials, it is thus necessary to
perform a direct analysis of the static domain wall configuration,
extracting the roughness exponent $\zeta$ and the effective domain
wall dimensionality $d_{\rm eff}$. Although measurements of this
kind have been performed on other elastic disordered systems such
as vortices (using neutron diffraction and decoration)
\cite{klein_brglass_nature,kim_decoration_nbse}, charge density
waves (using X-ray diffraction) \cite{rouzieres_structue_cdw},
contact lines \cite{moulinet_sed_wetting}, and ferromagnetic
domain walls \cite{lemerle_FMDW}, a successful comparison between
the experimentally observed roughness exponent and theoretical
predictions could only be carried out in magnetic systems. In
these systems, good agreement with the value $\zeta = 2/3$
predicted for one-dimensional (line) domain walls in a random bond
disorder was found \cite{lemerle_FMDW}. Quantitative studies of
these phenomena in other microscopic systems are therefore clearly
needed. Epitaxial perovskite ferroelectric thin films with high
crystalline quality and precisely controllable thickness are an
excellent model system for such studies.

In this paper we report on the first direct measurement of
ferroelectric domain wall roughness. To image ferroelectric
domains with the nanometer resolution required by such studies,
atomic force microscopy (AFM) was used
\cite{PP_afm_arrays,cho_AFM_Tbit,tybell_creep,PP_dw_dynamics_FE}.
Relaxation of the domain walls to their equilibrium configuration
at short length scales allowed us to obtain values of $\sim$ 0.26
for the roughness exponent $\zeta$. In the same films, the dynamic
exponent $\mu$ was found to be $\sim$ 0.5--0.6 from independent
measurements of domain wall creep. An analysis of these results
gives an effective dimensionality of $d \sim 2.5$ for the domain
walls, in good agreement with theoretical calculations for a
two-dimensional elastic interface in the presence of random-bond
disorder and long range dipolar interactions
\cite{nattermann_dipole_disorder}.

The ferroelectric domain wall roughness studies were carried out
in three c-axis oriented \PZT films, 50, 66 and 91 nm thick,
epitaxially grown on single crystalline (001) Nb:SrTiO$_3$
substrates by radio-frequency magnetron sputtering, as detailed in
ref.~\cite{tybell_FE_films,triscone_FE_films}. In these films, the
polarization vector is parallel or anti-parallel to the c-axis and
can be locally switched by the application of voltage signals via
a metallic AFM tip, using the conductive substrate as a ground
electrode \cite{guthner_afm_FE,ahn_afm_FE}. The resulting
ferroelectric domains are imaged by piezoforce microscopy (PFM)
\cite{guthner_afm_FE}. To measure domain wall roughness, we wrote
linear domain structures with alternating polarization by applying
alternating $\pm 12$ V signals while scanning the AFM tip in
contact with the film surface. We chose line widths of 1 -- 1.5
$\mu$m, and lengths of 8 -- 15 $\mu$m to ensure that domain wall
images ($2.5 \times 2.5$ and $5 \times 5$ $\mu$m$^2$) used in the
study could be taken in the central regions, away from possible
edge effects. Multiple domain structures were written in
photolithographically pre-defined areas on each sample. More than
100 different ferroelectric domain walls were written and imaged
in the three films.

\begin{figure}
\includegraphics{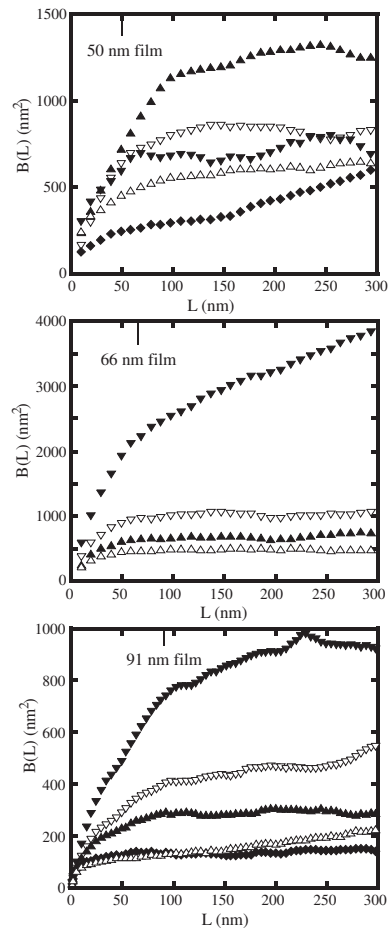}
\caption{Average displacement correlation function $B(L)$ for
different sets of ferroelectric domain walls in 50 (a), 66 (b) and
91 (c) nm thick films, shown out to $L = 300$ nm.  Power law
growth of $B(L)$ is observed at short length scales, followed by
saturation, suggesting a non-equilibrium configuration at large
$L$.} \label{B_L_all}
\end{figure}

From these measurements, we extracted the correlation function of
relative displacements
\begin{equation}\label{eq:B_gen}
 B(L)  = \overline{\langle [u(z+L)-u(z)]^2 \rangle}
\end{equation}
where the displacement vector $u(z)$ measures the deformation of
the domain wall from an elastically optimal flat configuration due
to pinning in favorable regions of the potential landscape.
$\langle\cdots\rangle$ and $\overline{\cdots}$ denote the
thermodynamic and ensemble disorder averages, respectively.
Experimentally, the latter is realized  by averaging over all
pairs of points separated by the fixed distance $L$, ranging from
1 to 500 pixels (5 or 10 nm -- 2.5 or 5.0 $\mu$m, depending on the
image size) in our measurements. As shown in \fref{B_L_all} for
the three different films used, we observe a power-law growth of
$B(L)$ at short length scales, comparable to the $\sim$ 50--100 nm
film thickness, followed by saturation of $B(L)$ in the 100--1000
nm$^2$ range \footnote{This saturation is apparent well below $L
\sim$ 1.25 or 2.5 $\mu$m, the limit at which the finite size of
the image begins to play a role.}.
The observed $B(L)$ saturation indicates that the walls do not
relax at large length scales from their initial straight
configuration, dictated by the position of the AFM tip during
writing. To ensure that domain wall relaxation was not hindered by
the pinning planes of the lattice potential in the ferroelectric
films \cite{poykko_periodic_FE_DW,meyer_periodic_FE_DW}, we wrote
sets of domain walls at different orientations with respect to the
crystalline axes in the 66 nm film. We found no correlation
between the roughness of domain walls and their orientation in the
crystal. This result is in agreement with previous studies
\cite{tybell_creep} pointing out the negligible role of the
commensurate potential compared to the effects of disorder.

To investigate the possibility of thermal relaxation at ambient
conditions, we then measured a set of domain walls over a period
of 1 month.
\begin{figure}
\includegraphics{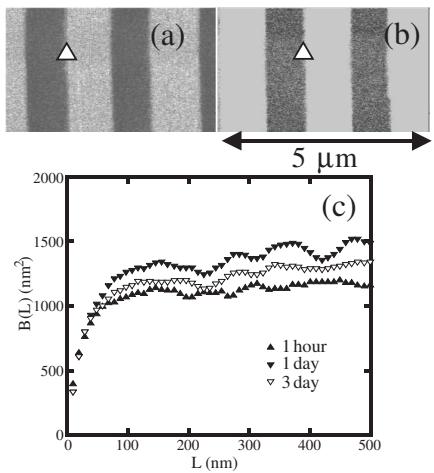}
\caption{PFM images of the same set of domain walls taken (a) 1
hour after writing and (b) 1 week later. The white triangle
indicates the same domain wall in each image. (c) The average
$B(L)$ for this set of domain walls.} \label{stability}
\end{figure}
No relaxation from the flat as-written configuration at large $L$
is apparent visually (\fref{stability}(a)), or when comparing
$B(L)$ extracted for this set of domain walls at different times
(\fref{stability}(b)). These data strongly indicate that ambient
thermal activation alone is not sufficient to equilibrate the
domain walls over their entire length \footnote{In other elastic
disordered systems relaxation was achieved by the applying small
oscillating driving forces
\cite{yaron_neutrons_vortex,joumard_shaking_vortex}, or by
subcritically driving the manifold into an equilibrium
configuration \cite{lemerle_FMDW}. Using Pt electrodes deposited
on top of the writing areas we therefore applied both DC and AC
voltages to different domain structures. Unfortunately, the large
degree of leakage in such thin films, avoided in the local probe
configuration, did not allow the domain walls to reach equilibrium
at large $L$, with no relaxation observed when comparing $B(L)$
for domain wall imaged before and after voltage application.}.
These results are in agreement with our previous studies
\cite{kumar_SAW,PP_dw_dynamics_FE} in which both linear and
nanoscopic circular domains remained completely stable over 1--5
month observation periods. Such high stability is inherent to the
physics of an elastic disordered system, where energy barriers
between different metastable states diverge as the electric field
driving domain wall motion goes to zero. This is an advantage for
possible memory or novel filter applications \cite{kumar_SAW}, but
also makes relaxation exceedingly slow. In fact, we believe that
even the relaxation leading to the observed power-law growth of
$B(L)$ at smaller length scales is not purely thermal, but occurs
during the writing process itself. When the direction of the
applied electric field is reversed to form the alternating domain
structure, the neighboring region already written with the
opposite polarity nonetheless experiences the resulting electric
field, allowing the domain wall to locally reach an equilibrium
configuration.

\begin{figure}
\includegraphics{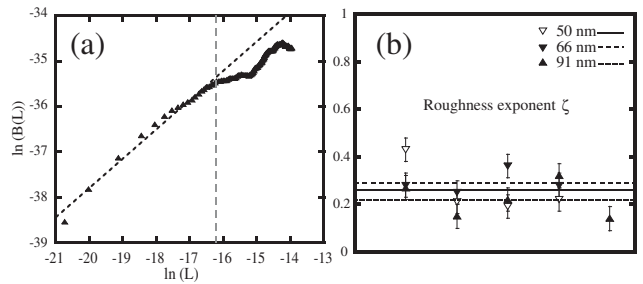}
\caption{(a) Typical ln-ln plot of $B(L)$.  Fitting the linear
part of the curve (left of the vertical line) gives 2$\zeta$. (b)
Average values of the characteristic roughness exponent $\zeta$
extracted from the equilibrium portion of the $B(L)$ data are
0.26, 0.29 and 0.22 in the 50, 66 and 91 nm thick samples,
respectively, indicated by the horizontal lines in the figure.}
\label{zeta}
\end{figure}

From the power-law growth of $B(L)$ at these short length scales,
we extract a value for the roughness exponent $\zeta$. This
exponent characterizes the roughness of the domain wall in the
random manifold regime where an interface individually optimizes
its energy with respect to the disorder potential landscape and
$B(L)$ scales as $B(L) \propto L^{2\zeta}$. As shown in
\fref{zeta}(a), a linear fit of the lower part of the $\ln(B(L))$
vs $\ln(L)$ curve allows $2\zeta$ to be determined. Average values
of $\zeta \sim$ 0.26, 0.29 and 0.22 were obtained for the 50, 66
and 91 nm thick films, respectively, indicated by the horizontal
lines in \fref{zeta}(b).

\begin{figure}
\includegraphics{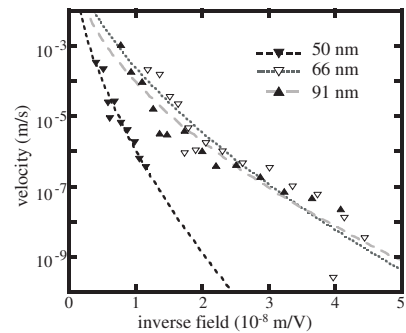}
\caption{Domain wall speed as a function of the inverse applied
electric field extracted from measurements of domain growth in the
50, 66 and 91 nm thick films. The data fit well to $v \sim
\exp[-C/E^\mu]$ with $\mu =$ 0.59, 0.58 and 0.51, respectively.}
\label{creep}
\end{figure}

In addition to the investigations of static domain wall roughness
described above, we independently measured domain wall dynamics in
each film, using the approach detailed in
\cite{tybell_creep,PP_dw_dynamics_FE}.  As shown in \fref{creep},
we observe the non-linear velocity response to applied electric
fields characteristic of a creep process, with values of 0.59,
0.58 and 0.51 for the dynamical exponent $\mu$ in the 50, 66 and
91 nm thick films, respectively \footnote{These $\mu$ values are
lower than the three values measured in \cite{tybell_creep}, but
consistent with all the subsequent measurements performed on nine
other films, all grown under similar conditions.}.

When these data are analyzed in the theoretical framework of a
disordered elastic system, they provide information on the
microscopic mechanism governing domain wall behavior. The direct
measurement of domain wall roughness clearly rules out the lattice
potential as a dominant source of pinning. In that case, the walls
would have been flat with $B(L) \sim a^2$, where the lattice
spacing $a \sim 4$ \AA~is the period of the pinning potential
\cite{meyer_periodic_FE_DW}. Given the stability and
reproducibility of the wall position over time, shown in
\fref{stability}, the effect of thermal relaxation on the observed
increase of $B(L)$ can also be ruled out.  The measured roughness
must thus be attributed to disorder. Two disorder universality
classes exist, with different roughness exponents. Random bond
disorder, corresponding to defects maintaining the symmetry of the
two polarization states would change only the local depth of the
ferroelectric double well potential. Theoretically, this disorder
would lead to a roughness exponent $\zeta_{RB} = 2/3$ in $\deff=1$
and $\zeta_{RB} \sim 0.2084(4-\deff)$ for other dimensions. Random
field disorder, corresponding to defects which locally asymmetrize
the ferroelectric double well would favor one polarization state
over the other. Such disorder would lead to a roughness exponent
$\zeta_{RF} =(4-\deff)/3$ in all dimensions below four. Should the
wall be described by standard (short range) elasticity, $\deff$ in
the above formulas is simply the dimension $d$ of the domain wall
($d=1$ for a line, $d=2$ for a sheet). However, in ferroelectrics
the stiffness of the domain walls and thus their elasticity under
deformations in the direction of polarization is different from
the one for deformations perpendicular to the direction of
polarization because of long range dipolar interactions
\cite{nattermann_dipole_disorder}. The elastic energy (expressed
in reciprocal space) thus contains not only a short range term $H
= \frac12 \sum_{q} C_{\rm el}(q) u^*(q) u(q)$ with $C_{\rm el} =
\sigma_w q^2$ but also a correction term due to the dipolar
interaction $ C_{\rm dp} =
\frac{2P_s^2}{\epsilon_0\epsilon}\frac{q_y^2} {q} +
\frac{P_s^2\xi}{\epsilon_0\epsilon}\left(\frac{-3}{4}q_x^2 +
\frac{1}{8}q^2\right)$ where $y$ is the direction of the
polarization. $P$ is the ferroelectric polarization and $\epsilon$
and $\epsilon_0$ are the relative and vacuum dielectric constants.
Because  $q_y$ now scales as $q_y \sim q_x^{3/2}$, the effective
dimension $d_{\rm eff}$ to use in the above formulas is $d_{\rm
eff} = d + 1/2$
\cite{nattermann_dipole_disorder,emig_commdisorder_long}. Using
the above expressions for the roughness exponent we see that the
measured $\zeta \sim 0.26$ value would give $d_{\rm eff} \geq 3$
for random field disorder, ruling out this scenario. On the other
hand random bond disorder would give $d_{\rm eff} \sim 2.5$ --
$2.9$, a much more satisfactory value, which is compatible with a
scenario of two-dimensional walls (sheets) in random bond disorder
with long range dipolar interactions.

This conclusion can be independently verified by the dynamic
measurements, since the creep exponent $\mu$ is related to the
static roughness exponent $\zeta$ via $\mu = \frac{d_{\rm eff} - 2
+ 2\zeta}{2 - \zeta}$. The values of these two exponents from the
independent static and dynamic measurements can therefore be used
to calculate $d_{\rm eff}$. For the 50, 66 and 91 nm thick films
we find $d_{\rm eff} = 2.42$, 2.49 and 2.47, respectively, in very
good agreement with the expected theoretical value for a
two-dimensional elastic interface in the presence of disorder and
dipolar interactions. Taken together, these two independent
analyses provide strong evidence that the pinning in thin
ferroelectric films is indeed due to disorder in the random bond
universality class. The precise microscopic origin of such
disorder is still to be determined. Preliminary studies of domain
wall dynamics in pure \PTO films gave comparable results to those
for PbZr$_{0.2}$Ti$_{0.8}$O$_3$, suggesting that the presence of
Zr in the solid solution is not a dominant factor, and that other
defects presumably play a more significant role. Note that for the
short-range domain wall relaxation observed, the walls are in the
two-dimensional limit. However, if equilibrium domain wall
roughness could be measured for larger $L$, a crossover to
one-dimensional behavior would be expected, with a roughness
exponent $\zeta = 2/3$.

In conclusion, we used AFM measurements of high quality epitaxial
\PZT thin films to obtain the roughness exponent $\zeta$ for
ferroelectric domain walls. This is the first direct observation
of static domain wall roughness in ferroelectric systems, and,
combined with our measurements of domain wall dynamics, provides a
coherent physical image of their behavior in the framework of
elastic disordered systems.

\begin{acknowledgments}
This work was supported by the Swiss National Science Foundation
through the National Center of Competence in Research ``Materials
with Novel Electronic Properties-MaNEP'' and Division II. Further
support was provided by the New Energy and Industrial Technology
Development Organization (NEDO) and the European Science
Foundation (THIOX).
\end{acknowledgments}

\bibliographystyle{prsty}

\end{document}